\documentclass[reprint, prd]{revtex4-1}

\usepackage[utf8]{inputenc}
\usepackage{amsmath}
\usepackage{amssymb}
\usepackage{mathrsfs}
\usepackage{times}
\usepackage{hyperref}
\usepackage{natbib}
\usepackage{graphicx}
\usepackage{hyperref}

\hypersetup{
  bookmarksnumbered = true,
  bookmarksopen=false,
  pdfborder=0 0 0,         
  pdffitwindow=true,      
  pdfnewwindow=true, 
  colorlinks=true,           
  linkcolor=blue,            
  citecolor=magenta,    
  filecolor=magenta,     
  urlcolor=cyan              
}

\newcommand{\nn}{\nonumber}

\newcommand{\chimera}{{\sc Chimera }}

\newcommand{\aap}{Astron. Astrophys.}
\newcommand{\mnras}{Mon. Not. R. Astron. Soc.}

\begin{document}

\title{Gravitational Wave Signatures of {\em Ab Initio} Two-Dimensional\\ Core Collapse Supernova Explosion Models for 12--25 M$_\odot$ Stars}

\author{Konstantin N. Yakunin$^{1,2,3}$,  Anthony Mezzacappa$^{1,2}$, Pedro Marronetti$^4$, Shin'ichirou Yoshida$^5$, Stephen W. Bruenn$^6$,
        W. Raphael Hix$^{1,3}$, \\ Eric J. Lentz$^{1,2,3}$, O. E. Bronson Messer$^{1,4,7}$, 
        J. Austin Harris$^1$, 
        Eirik Endeve$^{1,8}$, John M. Blondin$^9$,
        Eric J. Lingerfelt$^{3,8}$}

\affiliation{}        
\affiliation{$^1$Department of Physics and Astronomy, University of Tennessee, Knoxville, TN 37996-1200, USA}
\affiliation{$^2$Joint Institute for Computational Sciences, Oak Ridge National Laboratory, P.O. Box 2008, Oak Ridge, TN 37831-6354, USA}
\affiliation{$^3$Physics Division, Oak Ridge National Laboratory, P.O. Box 2008, Oak Ridge, TN 37831-6354, USA}
\affiliation{$^4$Physics Division, National Science Foundation, Arlington, VA 22230 USA}
\affiliation{$^5$Department of Earth Science and Astronomy, Graduate School of Arts and Sciences, The University of Tokyo, Komaba, Meguro-ku, Tokyo 153-8902, Japan}
\affiliation{$^6$Department of Physics, Florida Atlantic University, 777 Glades Road, Boca Raton, FL 33431-0991, USA}
\affiliation{$^7$National Center for Computational Sciences, Oak Ridge National Laboratory, P.O. Box 2008, Oak Ridge, TN 37831-6164, USA}
\affiliation{$^8$Computer Science and Mathematics Division, Oak Ridge National Laboratory, P.O.Box 2008, Oak Ridge, TN 37831-6164, USA}
\affiliation{$^9$Department of Physics, North Carolina State University,  Raleigh, NC 27695-8202, USA}

\email{kyakunin@utk.edu}

\begin{abstract}
We present the gravitational waveforms computed in \emph{ab initio} two-dimensional core collapse supernova models evolved with 
the \chimera\ code for progenitor masses between 12 and 25 M$_\odot$. 
All models employ multi-frequency neutrino transport in the ray-by-ray approximation, state-of-the-art weak interaction physics, relativistic transport corrections 
such as the gravitational redshift of neutrinos, two-dimensional hydrodynamics with the commensurate relativistic corrections, Newtonian self-gravity 
with a general relativistic monopole correction, and the Lattimer--Swesty equation of state with 220 MeV compressibility, and begin with 
the most recent Woosley--Heger nonrotating progenitors in this mass range.
All of our models exhibit robust explosions. Therefore, our waveforms capture all stages of supernova development:  
1) a relatively short and weak prompt signal, 
2) a quiescent stage, 3) a strong signal due to convection and SASI activity,  
4) termination of active accretion onto the proto-neutron star, and 5) a slowly increasing tail that reaches a saturation value.
Fourier decomposition shows that the gravitational wave signals 
we predict should be observable by AdvLIGO for Galactic events across the range of progenitors considered here. 
The fundamental limitation of these models is in their imposition of axisymmetry. 
Further progress will require counterpart three-dimensional models, which are underway.
\end{abstract}

\pacs{Valid PACS appear here}


\maketitle

\section{Introduction}
  
Core collapse supernovae (CCSNe) are recognized as the most energetic explosions in the modern Universe.
As a result of the collapse and the subsequent explosion of a massive star ($M > 9M_\odot$),
gravitational binding energy of the progenitor ($ 100\textrm{ B} = 10^{53}$ ergs) is released in the form of neutrino radiation ($ ~99$\%) and 
the kinetic energy of ejected material observed in the electromagnetic spectrum ($\sim 1$\%). 
Post-collapse asymmetries in fluid motion along with anisotropic emission of neutrinos also generate 
a strong burst of gravitational waves (GWs). The presence of three different types of radiation associated with
CCSNe makes them ideal objects for multi-messenger astronomy \cite{Ando13}. Simultaneous observations 
of the multi-messengers from CCSNe can reveal not only the details of the supernova mechanism \cite{Logue12}
but also may shed light on fundamental properties of neutrinos \cite{Cuesta08}. 
Gravitational radiation signals are particularly interesting among them because they are the only signals that provide 
deep insight into the multidimensional dynamics of the supernova core. 
The uniqueness of the information carried by GWs was recognized quite early 
(for comprehensive reviews of GWs from CCSNe, see \cite{Kotake07, Ott09, Kotake13}).

However, CCSNe are extremely complex and physically diverse phenomena that
involve an intricate interplay of general relativistic gravity, hydrodynamics, and neutrino transport, and thermonuclear kinetics, 
on short time scales. The complexity of the problem requires state-of-the-art numerical simulations for 
a quantitative analysis of the processes taking place in the supernova core. This is a great challenge in and of itself.
It is currently known that \emph{realistic} simulations of CCSNe are multidimensional and include hydrodynamics, self-gravity, and 
neutrino transport with the complete set of neutrino weak interactions, preferably in full general relativity but with at least approximate general relativistic corrections to all three.
Recent simulations of other groups that have provided gravitational waveforms met some of these requirements 
\cite{Murphy09, Mueller13, Ott13, Kuroda14}.

The first simulations of GW emission from core collapse supernovae were limited to rapidly rotating stellar core collapse \cite{Zwerger97, Dimmelmeier02, Kotake03, Ott04}. These simulations did not require a large amount of computational resources in order to produce the strongest part of the GW signal at bounce and, therefore, their primary waveforms are of relatively short duration ($\sim$30--50 ms). More recent predictions of the signal from rapidly rotating collapsing cores \cite{Ott07a, Dimmelmeier07a, Dimmelmeier08, Kotake09, Abdikamalov10, Kotake11, Takiwaki11, Ott12, Ott13, Abdikamalov14, Kuroda14, Fuller15} use two- or three-dimensional simulations, conformally-flat or full general relativity, along with, in some cases, a deleptonization prescription for the stellar core, developed to reproduce the results of a full neutrino transport treatment during the collapse phase \cite{Liebendoerfer05} and a neutrino leakage scheme for the post-bounce evolution \cite{Ott12}.  For the relatively short simulation time associated with these models, this approach is quite adequate. However, most supernova cores likely do not rotate rapidly \cite{Heger05} to produce a strong GW signal. 

Another class of CCSN simulations has focused on the GW signatures of hydrodynamic instabilities in the post-bounce phases -- in particular, convection inside the proto-neutron star (PNS), neutrino-driven convection in the post-shock region \cite{Bethe90, Herant94, Burrows95, Janka96, Mueller97}, and the standing accretion shock instability (SASI) \cite{Blondin03, Blondin06,  Ohnishi06, Foglizzo07, Scheck08,Iwakami09, Fernandez10}. These models require that we simulate at least several hundred milliseconds of post-bounce time during which time all instabilities develop and achieve their nonlinear state, and potentially saturate, and during which time explosion also develops. In this case, a far more realistic treatment of neutrino transport is crucial. 
Although there has been recent significant progress (e.g., see \cite{Kuroda15}), multidimensional CCSN models with realistic neutrino transport and full general relativity are not yet available. This is in part due to the shear computational cost associated with such models that cannot be paid on present-day supercomputing platforms. There are a few methods widely used to approximate neutrino transport in two- or three-dimensional models that do not require as much computational power:  a parameterized approximation for neutrino heating and cooling \cite{Mueller97, Kotake07, Murphy09, Kotake09},  gray neutrino transport schemes \cite{Fryer04, Mueller13}, the IDSA method \cite{Scheidegger10}, and finally, but most notably, multi-group neutrino transport (e.g., multi-group flux-limited diffusion) in the \textquotedblleft ray-by-ray-plus\textquotedblright~approximation \cite{Buras06}. 
Previous studies \cite{Murphy09, Marek09, Yakunin10, Mueller13} using one of these methods have established the fundamental structure of the waveform, which consists of three major components: a prompt-convection signal that lasts about 70~ms; the strongest part of the signal, due to neutrino-driven convection and the SASI ($\sim$200~ms); and a signal due to the revived shock's expansion, which produces a growing offset in GW amplitude. Uniqueness of the  \textquotedblleft ray-by-ray\textquotedblright~method is that it allows us to estimate the contribution of anisotropic neutrino emission to the total GW signal  \cite{Mueller04, Marek09, Yakunin10, Mueller13}.

In this paper, we present the gravitational waveforms from four non-rotating axisymmetric (two-dimensional) relativistic models evolved with the neutrino-hydrodynamics code \chimera beyond one second of post-bounce time and explosion.
We compare our results with both GW predictions from our previous simulations (A-series) \cite{Yakunin10} and the results of other groups \cite{Mueller13, Ott13}. We discuss all of the features of the GW signals from the four models initiated from four different progenitors with masses 12, 15, 20, and 25~M$_\odot$. All of our waveforms are available for download from \href{www.chimerasn.org}{ChimeraSN.org}.

Our paper is organized as follows. A brief description of the code and the model
setup is given in Section 2. In Section 3, we describe
the methods used for extraction and analysis of the GW signals. 
In Section 4, we provide gravitational waveforms
for all of our models and the results of our analysis.
In Section 5, we summarize our investigation and draw conclusions.
 \section{Code Description and Model Setup}
We analyze the GW emission in four two-dimensional simulations 
performed with the neutrino-hydrodynamics code \chimera~\cite{Bruenn14} . 
\chimera~consists of five major modules: hydrodynamics, neutrino transport,
self-gravity, a nuclear equation of state, and a nuclear reaction network. 

\chimera~solves the equations of Newtonian hydrodynamics but takes into account
some effects of strong-field gravity by means of an ``effective potential" \cite{Rampp02, Marek06}.
The gravitational field is computed by multipole expansion \cite{Mueller95}. We replace the Newtonian monopole component with a GR monopole \cite{Marek06}. 
It has been shown by \citet{Mueller10} that this approximation works very well for slow rotation.
The neutrino transport module solves the energy-dependent neutrino 
moment equations for all neutrino flavors using an updated version of multi-group
flux-limited diffusion (in the ray-by-ray-plus approach of \cite{Buras06}) with a flux limiter
that has been tuned to reproduce Boltzmann transport results \cite{Liebendoerfer04}. 
The ray-by-ray-plus method is able to produce angular variations in the 
neutrino radiation field, which, in turn, generates low frequency GW signals 
(Figure~\ref{fig:allmodelshplus}).

In total, we evolve four different non-rotating, non-perturbed, axisymmetric models (designated B12-WH07, B15- WH07, B20-WH07, and B25-WH07, corresponding to zero-age main sequence progenitors of 12, 15, 20, and 25 M$_\odot$ \cite{Woosley07}) on a spherical-polar mesh consisting of 512 non-equally-spaced, adaptive radial zones and 256 uniformly-spaced angular zones. In radius, the grid covers $3\times 10^4$, $2\times 10^4$, $2.1\times 10^4$, and $2.3\times 10^4$ km, respectively. In angle, the grid goes from 0 to $\pi$.

All presented models were simulated using the Lattimer--Swesty \cite{Lattimer91} equation
of state (EoS) with a bulk incompressibility modulus $K = 220$~MeV for $\rho > 10^{11} \textrm{g}\,\textrm{cm}^{-3}$, which is capable of supporting
the maximum observed neutron star masses of $\sim$2~M$_\odot$  \cite{Demorest10, Antoniadis13},
and an enhanced version of the Cooperstein EoS \cite{Cooperstein85} for $\rho < 10^{11} \textrm{g}\,\textrm{cm}^{-3}$ where nuclear statistical equilibrium (NSE) applies. 

All four progenitors are evolved beyond 1 second after core bounce. All models exhibit shock revival and the development of neutrino-driven explosions. The evolution beyond 1 second captures all important phases of the supernova dynamics that pronounce in the GW signals.  Three of the models developed clear prolate shock morphologies, while the 20~M$_\odot$ model develops an approximately spherical, off-center shock as the explosion begins, and then becomes moderately prolate at $\sim$600~ms after bounce. The morphologies of the explosions are reflected in the gravitational waveforms.
The explosion geometry of the 20~M$_\odot$  model has reduced not only the model's explosion energy relative to the 15 and 25~M$_\odot$  models but also the total energy emitted in the form of GWs.
%
%
 \section{Gravitational Wave Extraction}
 We consider the slow-motion, weak-field approximation
for post-processing extraction of the GW signal from our simulations.

\subsection{GWs Produced by a Time-Dependent, Mass-Quadrupole Moment}

We consider only the lowest order terms
in the retarded expansion of the mass-quadrupole formula \cite{Epstein75}.
The transverse-traceless part (TT) of the gravitational strain is given by
\begin{equation}
 h_{ij}^{TT} = \frac{1}{D}\sum_{m=-2}^{m=2}
  \left(\frac{d}{dt}\right)^2 I_{2m}\left(t-\frac{D}{c}\right)f^{2m}_{ij},
  \label{hTT_2}
\end{equation}
where $D$ is the distance from the source to the observer and the mass quadrupole $I_{2m}\left(t-\frac{D}{c}\right)$ is defined as

\begin{equation}
 I_{2m} = \frac{16\pi G}{5c^4}\sqrt{3}\int \tau_{00} Y^*_{2m}r^2dV.
\end{equation}
Here $\tau_{00}$ is the $(00)$-component of the linearized stress-energy tensor \footnote{Originally, the stress-energy tensor contains not only the matter contribution
but also the gravitational pseudo-tensor, which can be omitted in the linearized theory.}.
In the weak-field limit,  $\tau_{00}$ is approximated by the rest-mass density of matter.
The tensor spherical harmonics $f_{\ell m}(\theta,\phi)$ ($\theta$ and $\phi$ are the angular coordinates of the observer's frame ($O$-frame))
are defined in the Appendix.
The amplitude $A_{2m}$ in the gravitational strain is

\begin{equation}
 A_{2m} = \cfrac{d^2 I_{2m}}{dt^2}.
\end{equation}
Optimal design of gravitational wave detectors requires some knowledge of the expected waveforms,
the corresponding frequency spectra, and the total energy emitted by possible sources \cite{Hawking}.
Thus, any extraction method should decrease the numerical noise as much as possible.
Most numerical differentiation methods amplify the numerical noise built into the simulation data.
To avoid this, $A_{2m}$ is usually computed by reducing the order of time derivatives of $I_{2m}$:

\begin{equation}
 A_{2m} = \cfrac{dN_{2m}}{dt},\phantom{aaa}\textrm{where}\phantom{aaa} N_{2m} = \cfrac{dI_{2m}}{dt}.
\end{equation}
Following \cite{Nakamura89, Blanchet90} and \cite{Finn90}, the quadrupole signal 
can be expressed in terms of a volume integral
depending only on the density, velocity, and the gradient of the gravitational potential:
 
\begin{equation}
 \frac{d}{dt}\int \rho r^2Y^*_{2m} dV 
= \int \frac{\partial\rho}{\partial t} r^2Y^*_{2m} dV.
  \label{reduced_der}
\end{equation}
Using the continuity equation in the integrand in Eq.~(\ref{reduced_der}), one can replace
the density time derivative. Integrating by parts and omitting the surface contribution, 
we find the resulting integrand

\begin{eqnarray}
 N_{2m}& = &\cfrac{16\pi\sqrt{3}G}{c^4}
 \int_0^{2\pi}d\varphi' \int_0^\pi d\vartheta' \int_0^\infty dr'~r'^3 \\
  && \left[
2\rho v^{\hat{r}}Y^*_{2m}\sin\vartheta'
+ \rho v^{\hat{\vartheta}}\sin\vartheta'
\frac{\partial}{\partial\vartheta'}Y^*_{2m}
+ \rho v^{\hat{\varphi}}
\frac{\partial}{\partial\varphi'}Y^*_{2m}
\right] \nn
\label{n2m-integration}
\end{eqnarray}
where $r$, $\vartheta$ and $\varphi$ are the spherical coordinates in the source frame ($S$-frame), $v^{\hat{a}}$ is the component of velocity in the same frame.

Further reduction of the time derivative is conventionally 
done by using the momentum equation \cite{Nakamura89, Blanchet90}. 
However, this has to be carried out carefully.
The Euler equation includes stress terms. Therefore, in order to replace 
$\partial_t(\rho v)$ with balancing stress terms,
we need to take into consideration all possible contributions to the stress terms, such as
pressure, gravity, anisotropic neutrino forces acting on the fluid,
 effective viscosity that may present in the finite differencing of the momentum
equation, etc.
To avoid this issue, we have decided to compute
$N_{2m}$ for all time steps and to numerically evaluate
$d N_{2m}/dt$ to obtain $A_{2m}$. 
Numerical algorithms for computing first-order derivatives introduce far less numerical noise than 
those for higher-order derivatives. 
In order to determine the detectability of the GWs, 
we calculate the characteristic GW strain for a given frequency $f$ \cite{Flanagan98} using
\begin{eqnarray}
  h_c\left( f \right)= \frac{1}{D}\sqrt{\frac{2}{\pi^2}\,\frac{G}{c^3}\,\frac{dE_{GW}\left(f\right)}{df}}
\end{eqnarray}
\begin{eqnarray} 
    \frac{dE_{GW}\left(f\right)}{df}=
     \frac{c^3}{G}\,\frac{\left(2\pi f\right)^2}{16\pi} \left|\tilde{A}_{20}\left(f\right)\right|^2, \nonumber
\end{eqnarray}
where $dE_{GW}\left(f\right)/df$ is the GW energy spectrum and 
$\tilde{A}_{20}\left(f\right)$ denotes the Fourier transform of $A_{20}\left(t\right)$:
\begin{equation}
  \tilde{A}_{20}\left(f\right) = \int\limits_{-\infty}^{\infty}A_{20}\left(t\right)e^{-i\,2\pi ft}dt.
\end{equation}
The stochastic nature of GW signals from CCSNe prompts the use of short-time Fourier transform (STFT)
techniques to determine the frequency of a signal as it changes over time \cite{Murphy09}:
\begin{equation}
  \textrm{STFT}\{A_{20}(t)\}\left(\tau, f\right) =  \int\limits_{-\infty}^{\infty}
            A_{20}(t)\,H(t - \tau)e^{-i\,2\pi ft}dt
\end{equation}
where $H(t - \tau)$ is the Hann window function \cite{Blackman59}.


\subsection{Gravitational Waves Produced by Anisotropic Neutrino Emission}

Besides aspherical mass motion, any other sources with non-zero quadrupole moments
will produce GW emission.
One of these sources is the anisotropic radiation of neutrinos from the hot PNS. 
The theoretical derivation of the GW signal produced by a distant anisotropic point source of neutrinos 
was first published by \cite{Epstein78}.
M{\"u}ller end Janka \cite{Mueller97} were the first authors to implement this formalism. 
\citet{Kotake07} improved the formalism and made it more suitable for numerical evaluation of 
GW signals.

The transverse-traceless part of the gravitational strain, $h_{ij}^{TT}$, from neutrinos 
is given by  \cite{Mueller97}

\begin{equation}
 h_{ij}^\textrm{TT} = \frac{4G}{c^4 D}\int_{-\infty}^{t-D/c} dt'
  \int_\Omega d\Omega'~\frac{(n_in_j)^{TT}}{1-\cos\theta}
                   \,\frac{dL}{d\Omega'}(\vartheta',\varphi',t').
\label{h_nu_eq}
\end{equation}
Here $\Omega$ is the solid angle in the $S$-frame, and 
the vector $n_i$ is the direction of neutrino emission whose components are given with respect to the $O$-frame. 
The tensor $(n_in_j)^{TT}$ is the transverse-traceless part of the second-rank symmetric tensor $n_in_j$
with respect to the observer's $z$-axis (the $z$-direction is defined as the direction connecting the source and the observer). 
The angles $\theta$ and $\phi$ define the direction of neutrino emission with
respect to the $O$-frame. The other factor in the integrand,
$ dL /d\Omega$, is the \textquotedblleft direction-dependent neutrino
luminosity\textquotedblright~given in the $S$-frame.

In the case of axisymmetry, both $h_+$ and $h_\times$ components of the gravitational wave
signal vanish for an observer on the symmetry axis, and
the GW signal with the maximum amplitude will be detected by 
an observer in the equatorial plane \cite{Mueller97}. 
For such an observer, the gravitational strain is given by

\begin{equation}
 h_{\nu}^\textrm{TT} = \frac{2G}{c^4 D}\int_{-\infty}^{t-D/c} dt'
  \int_{4\pi} d\Omega'~\Psi(\vartheta',\varphi')\frac{dL}{d\Omega'}(\vartheta',\varphi',t'),
\label{h_nu}
\end{equation}
where

\begin{equation}
  \Psi\left(\vartheta',\varphi'\right) = \left( 1 + \sin\vartheta'\cos\varphi'\right)
     \frac{\cos^2\vartheta' - \sin^2\vartheta'\sin^2\varphi'}{\cos^2\vartheta' + \sin^2\vartheta'\sin^2\varphi'}.
\end{equation}
Since our models are axisymmetric, we can simplify  $\Psi\left(\vartheta',\varphi'\right)$. Integrating over $\varphi'$ \cite{Kotake07}:

\begin{eqnarray}
  \Psi\left(\vartheta'\right) & = & \sin\vartheta'\left( -\pi + \int_0^{2\pi} d\varphi'
     \frac{1 + \sin\vartheta'\cos\varphi'}{1 + \tan^2\vartheta'\sin^2\varphi'} \right) \nn \\
     & = & \pi\sin\vartheta'\left( -1 + 2\left|\cos\vartheta'\right|\right).
   \label{psi_function_eq}
\end{eqnarray}

\begin{figure*}
\centering
\includegraphics[scale = 0.8]{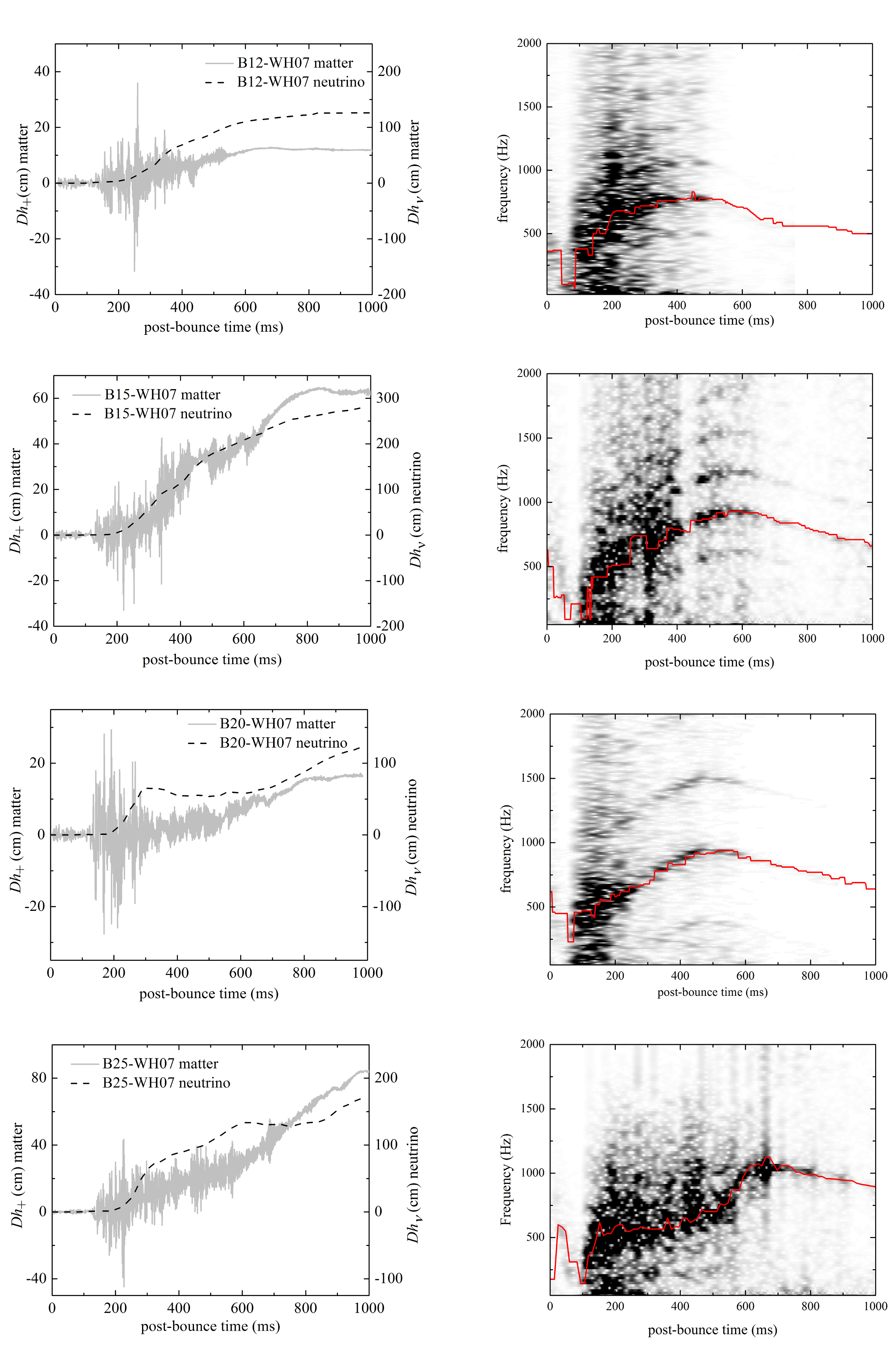}
\caption{\emph{Left panel:} The GW signals generated by both matter (solid gray) and neutrinos (dashed black) for our 12--25~M$_\odot$ models. \emph{Right panel:} Spectrogram showing a normalized value of $dE_\textrm{GW}^\textrm{matter}/df$ as a function of frequency and time after bounce. The blue line tracks the evolution of the peak frequency of the signal.}
\label{fig:allmodelshplus}
\end{figure*}   
%
%
\section{Waveform Analysis}
Gravitational waveforms covering all phases of GW emission and based on
non-parameterized supernova explosions were first reported in the work of \cite{Yakunin10}.
Here we present waveforms obtained in a new series of supernova simulations performed by
our group \cite{Bruenn13}, and a comparison with our previous results.

\subsection{Qualitative Description of the GW Signals}
 GW signals from both matter and neutrino sources for the models B12-WH07, B15-WH07, B20-WH07, and B25-WH07
 are provided in the left panels of Figure~\ref{fig:allmodelshplus}, 
and the evolution of the signal frequency  is given in the right panels of the same figure.
Qualitatively, our gravitational waveforms have all of the key features of waveforms
described in previous studies of CCSNe based on models that explode 
\cite{Marek09, Murphy09, Yakunin10, Mueller13}.
The GW signal passes through four distinct phases: 1) A prompt
signal: an initial and relatively weak signal that starts at bounce and ends at about 80 ms post-bounce. 
2) A quiescent stage that immediately follows the prompt
signal and ends somewhere about 125 ms after bounce. 3) A strong
signal, which follows the quiescent stage and is the most energetic part of the GW
signal. This stage ends somewhere between 350 ms and 400 ms after bounce. 4) 
A  ``tail," which starts just before the end of the strong phase, at about 300 ms after bounce and
consists of a low-frequency component with increasing amplitude. This ``tail" tends to rise during the simulations, 
but not monotonically. The signal produced by anisotropic neutrino emission exhibits a low-frequency ``tail"-like behavior during the entire simulation.

The frequency evolution in Figure~\ref{fig:allmodelshplus} shows how the characteristic frequency of the 
signal changes during the simulation. The initial frequency $\sim$500~Hz of the early signal drops to 100--200~Hz (the quiescent stage). The strong phase of the signal is characterized by a steady increase of the peak frequency, which reaches its maximum value of $\sim$700--800~Hz at 500--600 ms. Then, the frequency slowly decreases during the ``tail" phase.

\subsection{Early Signal}

\begin{figure}[ht!]
\centering
\includegraphics[width=86mm]{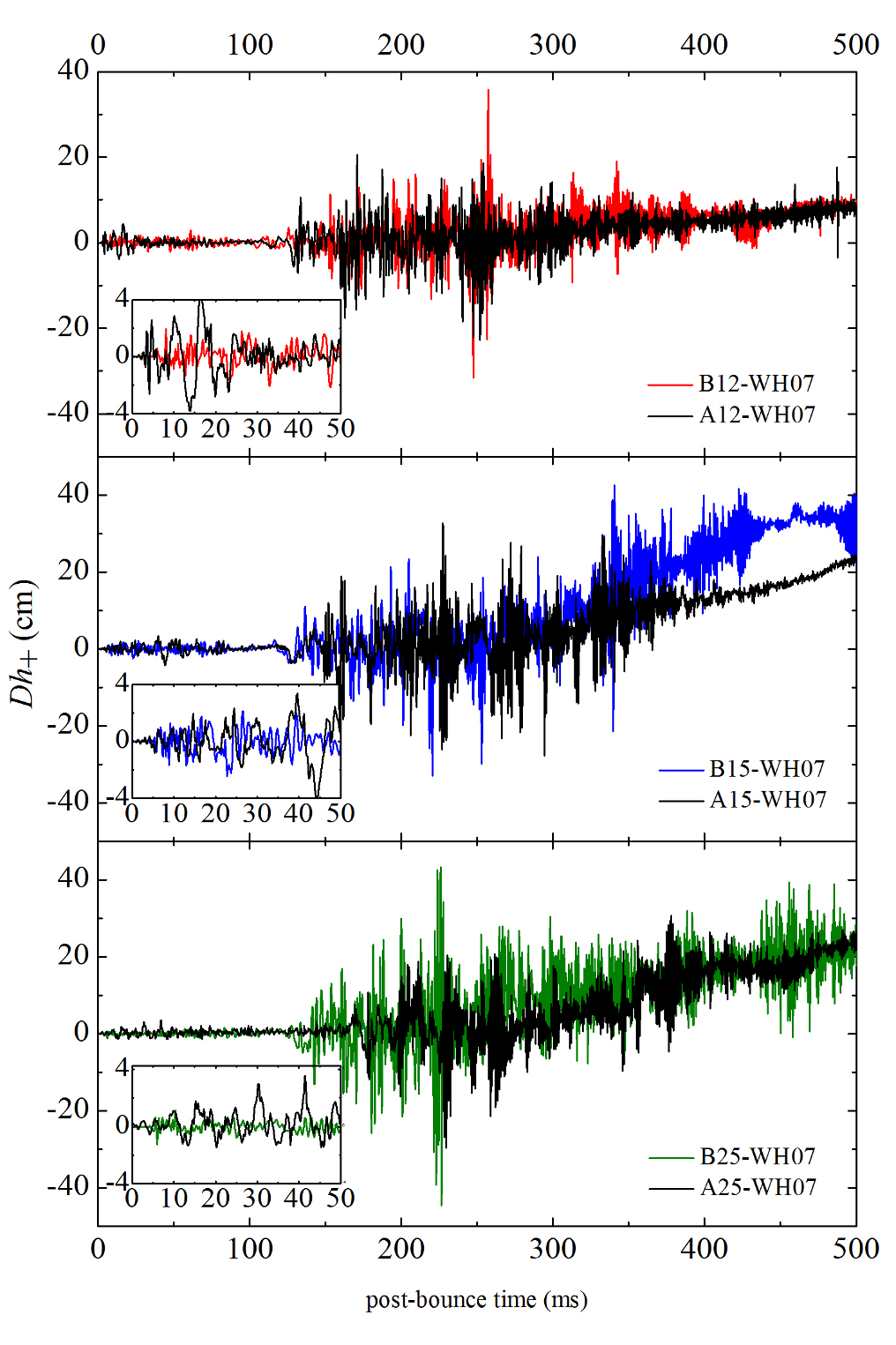}
\caption{Comparison of the waveforms  obtained in our A- and B-series \chimera simulations for the 12, 15 and 25 M$_\odot$ models.}
\label{fig:AvsB}
\end{figure}

The early GW signal is produced by entropy-driven convection in the PNS during the first 10 ms after bounce.
Unlike in our previously published models (the ``A-series") \cite{Yakunin10}, in the models presented here (the ``B-series")\cite{Bruenn13} we do not observe the
high-amplitude, low-frequency contribution to the prompt signal due to the deflection of infalling matter through the shock.
As a result, the peak amplitudes of the prompt signals in all the B-models decrease by a factor of 2--4 
relative to their values in the corresponding  A-series models (see the insets of Figure~\ref{fig:AvsB}).
The results presented here are in agreement with the conclusions drawn by \cite{Mueller13}, who 
estimated the contribution at the shock in their semi-analytic approach under the assumption that the 
non-radial component of the velocity is negligible. For our A-series runs, this assumption
was not valid. In that series, we set the lateral velocities of the fluid to zero above the shock 
during a short time after bounce for numerical reasons, which led to a coherent deflection of collapsing matter when 
passing through shock, with a sudden and commensurate change in the $\theta$-velocities of 
the fluid elements. One can see in Figure~\ref{fig:AvsB} the effect is more pronounced for a lesser progenitor mass. 
This happens because the relative contribution of the lateral velocity to the total velocity of the fluid is higher for less-massive progenitors 
(and, therefore, with lower radial velocities). 
In the B-series models, asymmetries in the angular velocities above 
the shock are obtained given commensurate asymmetries in the gravitational potential. This, 
in turn, reduces the jump in the angular velocities across the shock and excludes the presence of 
the coherent deflections seen in the A-series models and described above, as illustrated in Figure~\ref{fig:AvsB}. 

As was correctly pointed out in \cite{Mueller13}, prompt convection cannot be the only
source of the quasi-periodic signal in this emission phase, due to the difference in time scales of prompt convection 
($\sim$30 ms) and the prompt signal ($\sim$80 ms). After prompt convection has ceased, we do observe 
a diminishing but nonzero signal amplitude. The reason is that the shock expands quickly during the first 80 ms after bounce. 
Therefore, the perturbed matter is accumulated in the fast-growing volume behind the shock. Although convection has become 
less pronounced after 20 ms, integration over an increasing volume in part compensates for this. 

Further clarification of the contributions to the prompt GW signal would require simulations with a finer grid resolution that 
would be capable of capturing all hydrodynamic instabilities (i.e. prompt convection, turbulent cascades, acoustic and vorticity 
waves, etc.) that are presented in this phase.

The stalling of the shock manifests itself in a relatively short quiescent phase, which lasts about 20--30 ms and which is followed by 
a strong signal produced by the development of neutrino-driven convection and the SASI.

\subsection{Strong Signal Phase}

\begin{figure}[ht!]
\centering
\includegraphics[width=90mm]{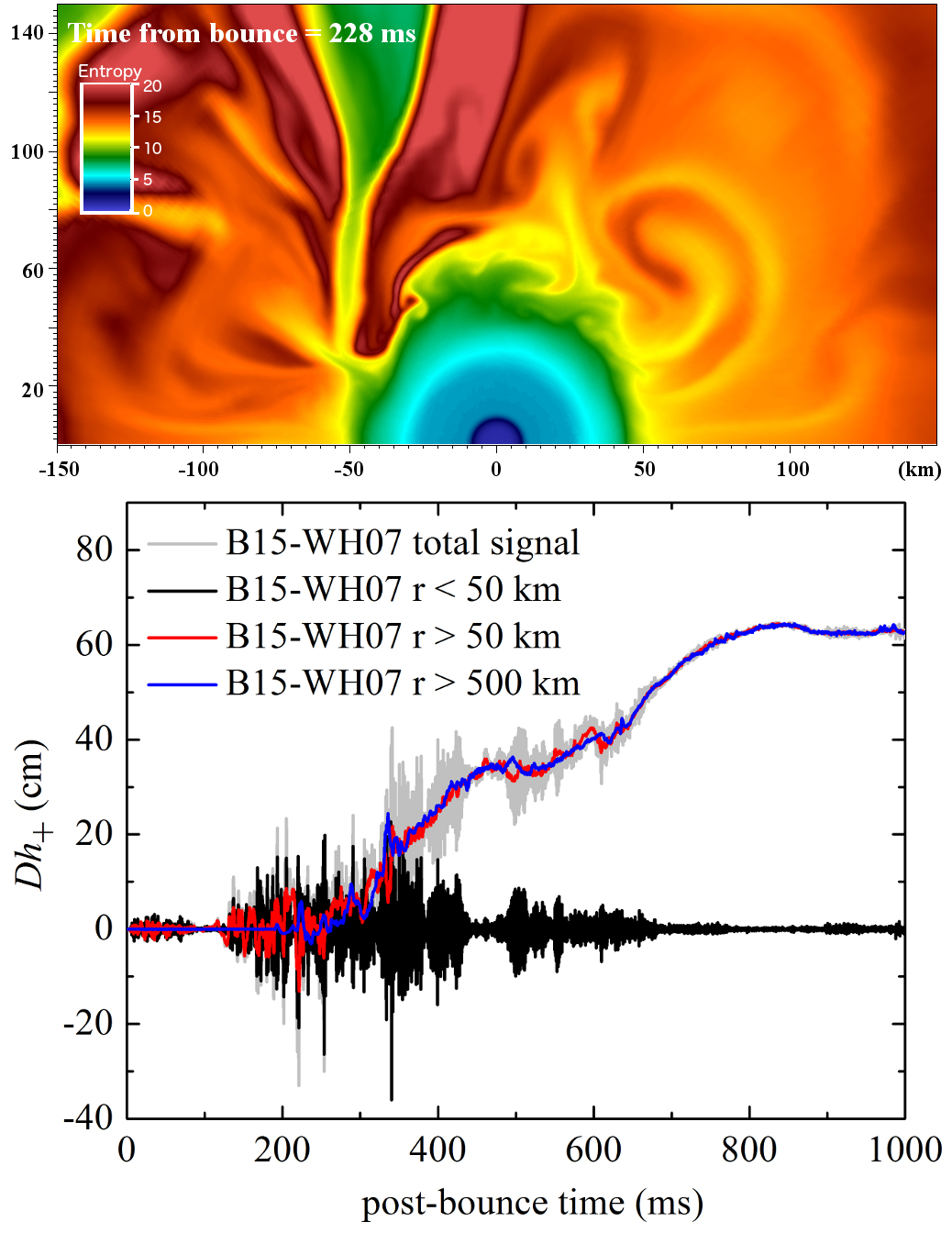}
\caption{
\emph{Top: } The entropy distribution for the B15-WH07 model inside the PNS at 228 ms after bounce. Down flows onto and convective activity inside the high-density region produce the strongest GW signal. \\
\emph{Bottom: } The GW waveforms, $Dh_+$ vs. time, showing the contributions of three regions: $r<$ 50~km, $r  >$ 50~km and $r >$ 500~km. The latter region shows the contribution due to the shock expansion.}
\label{fig:B15Decomp}
\end{figure}

The beginning of the strong signal phase coincides with the onset of SASI activity.
It has been shown in previous studies \cite{Murphy09, Marek09, Yakunin10} that
the strong signal is actually produced by the combined effect of
SASI-induced downflows impinging on the PNS surface and the subsequent deceleration of the matter at the PNS surface and the convection inside the PNS.
The low-frequency component arises from the modulations in the shock radius as the SASI develops and evolves. 
The high-frequency component is generated when the SASI-induced accretion flows strike
the PNS surface (Figure~\ref{fig:B15Decomp}). It is clear from the analysis of the contributions to the strain from $r<50$ km and $r>50$ km that the PNS convection, deceleration of the accreting matter at the PNS surface, and neutrino-driven convection in the gain region contribute significantly. 

The shock modulations affect the kinetic energy of the accretion
flows and, consequently, the amplitude of the GWs generated when these flows hit the
PNS surface. The signal structure during the strong signal phase in both B12-WH07 and B15-WH07 is similar to that in the corresponding A-series models. However, this is not the case for B25-WH07 and A25-WH07. The beginning of the strong signal phase in A25-WH07 is $\sim$50~ms behind that in B25-WH07, which indicates an earlier development of neutrino-driven convection and SASI activity in the latter model.  The peak amplitude in B25-WH07 is twice as large as it is in A25-WH07. 

The peak frequency of the signal grows almost linearly from 100~Hz up to 1000~Hz during the strong signal phase (right panels of Figure~\ref{fig:allmodelshplus}). We see the same trend in frequency evolution, with a similar slope, in the M15 model from \citet{Mueller13}, which is the closest to our B15-WH07 model.

\subsection{Explosion Phase}

\begin{figure}[ht!]
\centering
\includegraphics[width=90mm]{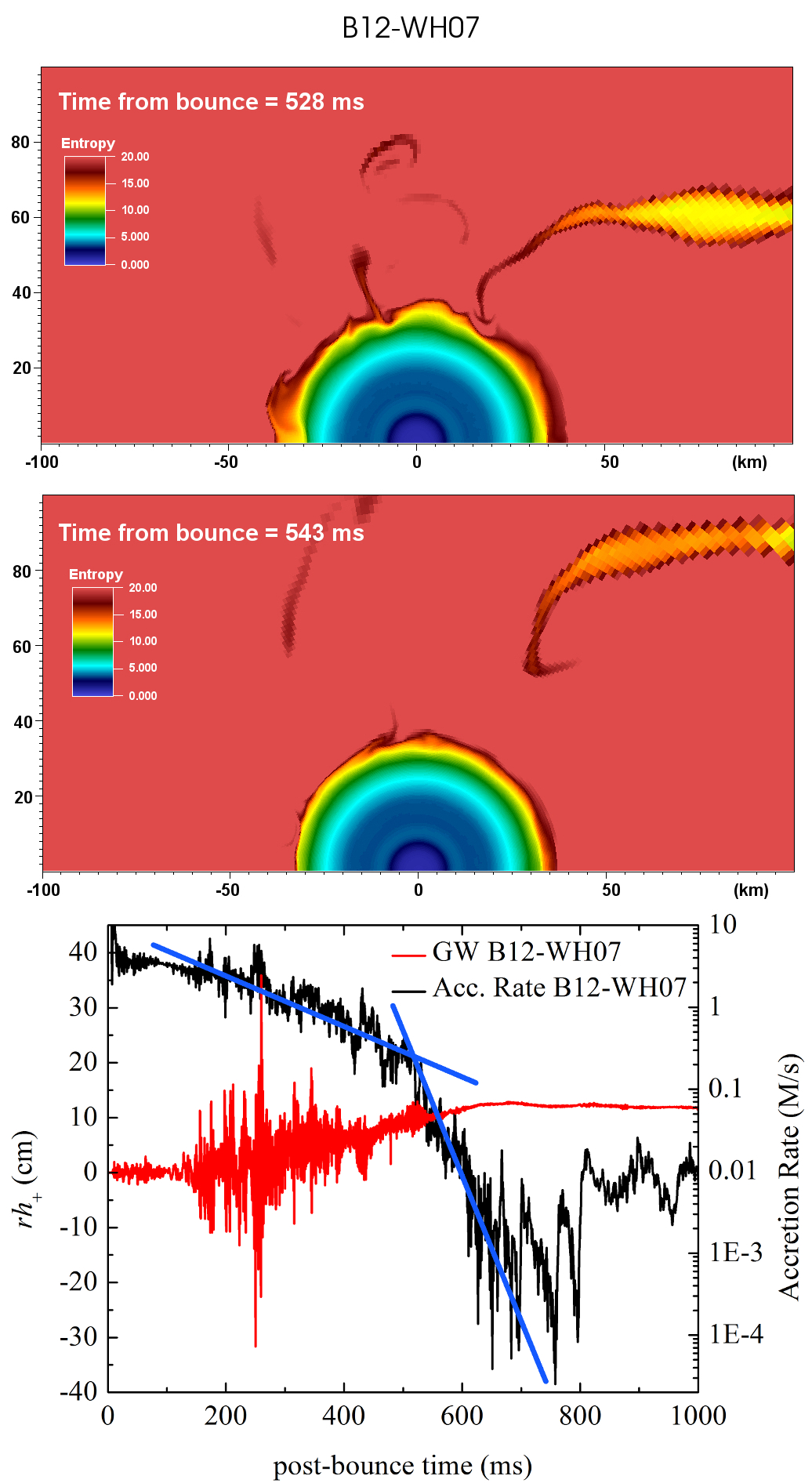}
\caption{The entropy distributions for the B12-WH07 model are shown at two post-bounce times: just before the accretion downstream finally detaches from the surface of the PNS (\emph{top panel}) and just after the detachment (\emph{middle panel}). This detachment produces a clear signature in the gravitational waveform. \\
\emph{Bottom panel: } The amplitude of the high-frequency component of the GW signal sharply decreases at the time ($\sim$530~ms) when the active accretion rate onto the PNS surface ( $\sim$50~km) drops significantly (right scale).  Further variations of the amplitude of the signal are mainly the result of the interaction between the outgoing shock and the infalling matter. The blue lines locate the significant change in the rate of change of the accretion rate, at $\sim$450~ms.}
\label{fig:B12Detach}
\end{figure}

\begin{figure}[ht!]
\centering
\includegraphics[width=86mm]{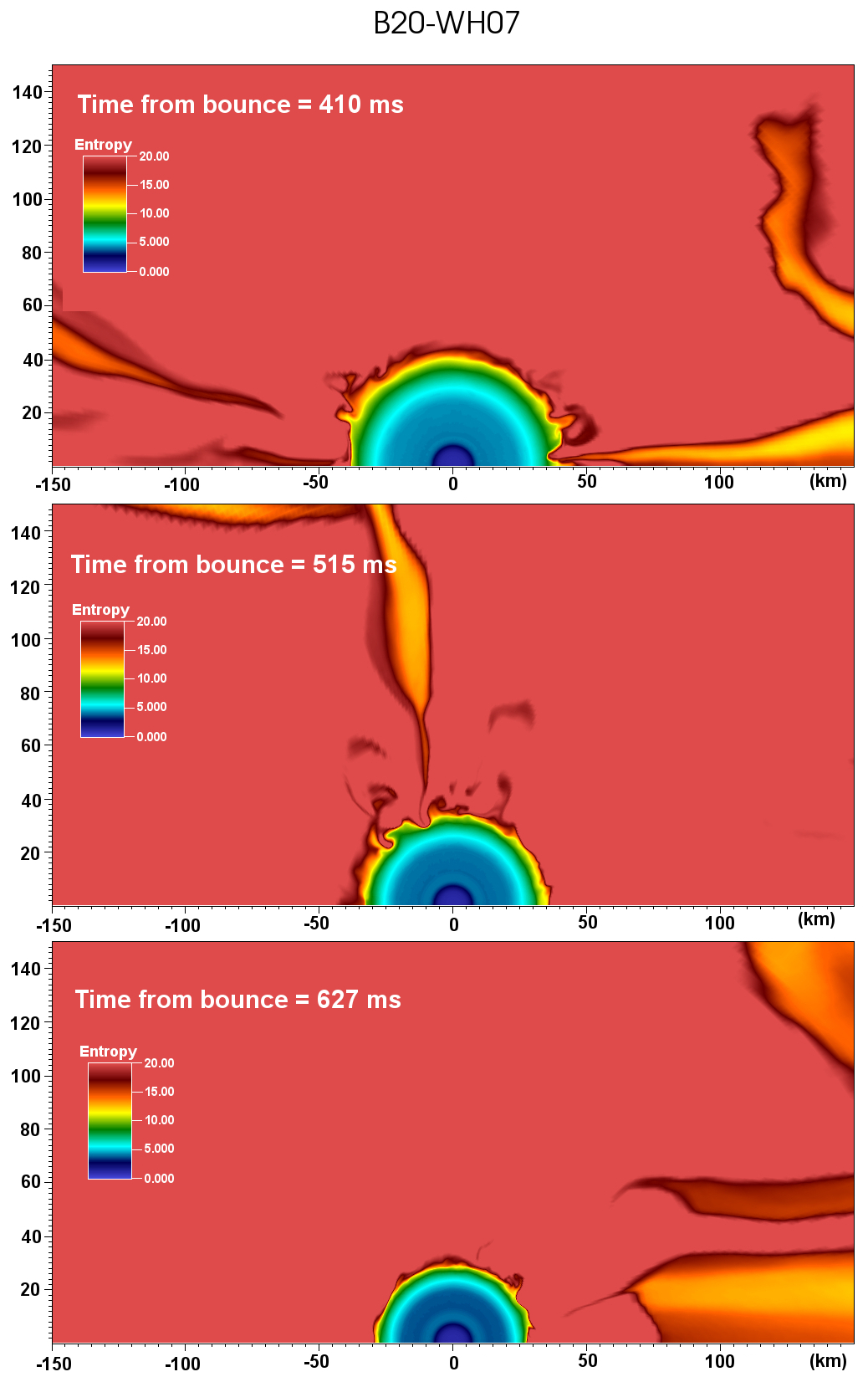}
\caption{Entropy distribution snapshots of the B20-WH07 model that show multiple down streams observed during the early accretion phase (\emph{top panel});
a single accretion down stream at a significantly later post-bounce time (\emph{middle panel});
the moment of final detachment of the final accretion down flow from the surface of the PNS (\emph{bottom panel}). 
This explains why the time when the peak frequency of the signal reaches its maximum ($\sim$500~ms) doesn't correlate well with the time when active accretion onto the PNS finally ceases. In B20-WH07, we begin with multiple accretion streams. The peak frequency begins to decline as the number of accretion streams is reduced, not as the single accretion stream detaches from the PNS, as in the other models. Note also: The significant change in the compactness of the PNS due to neutrino radiation emphasizes the importance of neutrino transport and general relativity in supernova simulations and to the accurate prediction of the associated gravitational waveforms.}
\label{fig:B20Detach}
\end{figure}

All of our GW signals end with a slowly increasing tail, which reflects the
(linear) gravitational memory associated with accelerations at the prolate outgoing shock.
The noticeable decrease of the high-frequency component of the amplitude during the explosion phase 
(most pronounced for model B12-WH07 at 520 ms) is due to the cessation of active accretion onto the PNS surface
(Figure~\ref{fig:B12Detach}). The time of the cessation coincides, within a width of the STFT window, with the time when the frequency reaches its maximum value, for all of our models except B20-WH07 (Figure~\ref{fig:allmodelshplus}). B20-WH07 has a different explosion morphology. A single downstream is formed in all of our models except B20-WH07 in the early SASI phase. This downstream produces the local large amplitude spikes in the GW strain by its deceleration at the PNS surface. 
The downflow also induces the \emph{l=2} mode of the mass distribution deep in the PNS, which enables high-frequency PNS convection to contribute to the GW signal. 
Thus, PNS convection is responsible for the high-frequency component of the GW waveform. Termination of the single accretion stream leads to a significant decrease in both the frequency and the amplitude of the GW signal. In B20-WH07, multiple downstreams are formed during the SASI phase. This prevents the establishment of a more precise correlation between the changes of the accretion flow and the associated changes of the waveform amplitude and peak frequency. The typical frequency in B20-WH07 starts to decrease when the first accretion downflow detaches from the surface of the PNS ($\sim$500~ms) while other downstreams continue to perturb the PNS and thus support the high-frequency and the amplitude of the B20-WH07 signal (Figure~\ref{fig:B20Detach}), until the moment when the last accretion down flow becomes detached from the PNS surface ($\sim$630~ms).
After the cessation of accretion, the GW signal in all of our models is essentially generated by the shock only.
The tails continue to rise until they reach their saturation values at 700--1000 ms, depending on 
the model and its prolateness.
 
\begin{figure}[ht!]
\centering
\includegraphics[width=86mm]{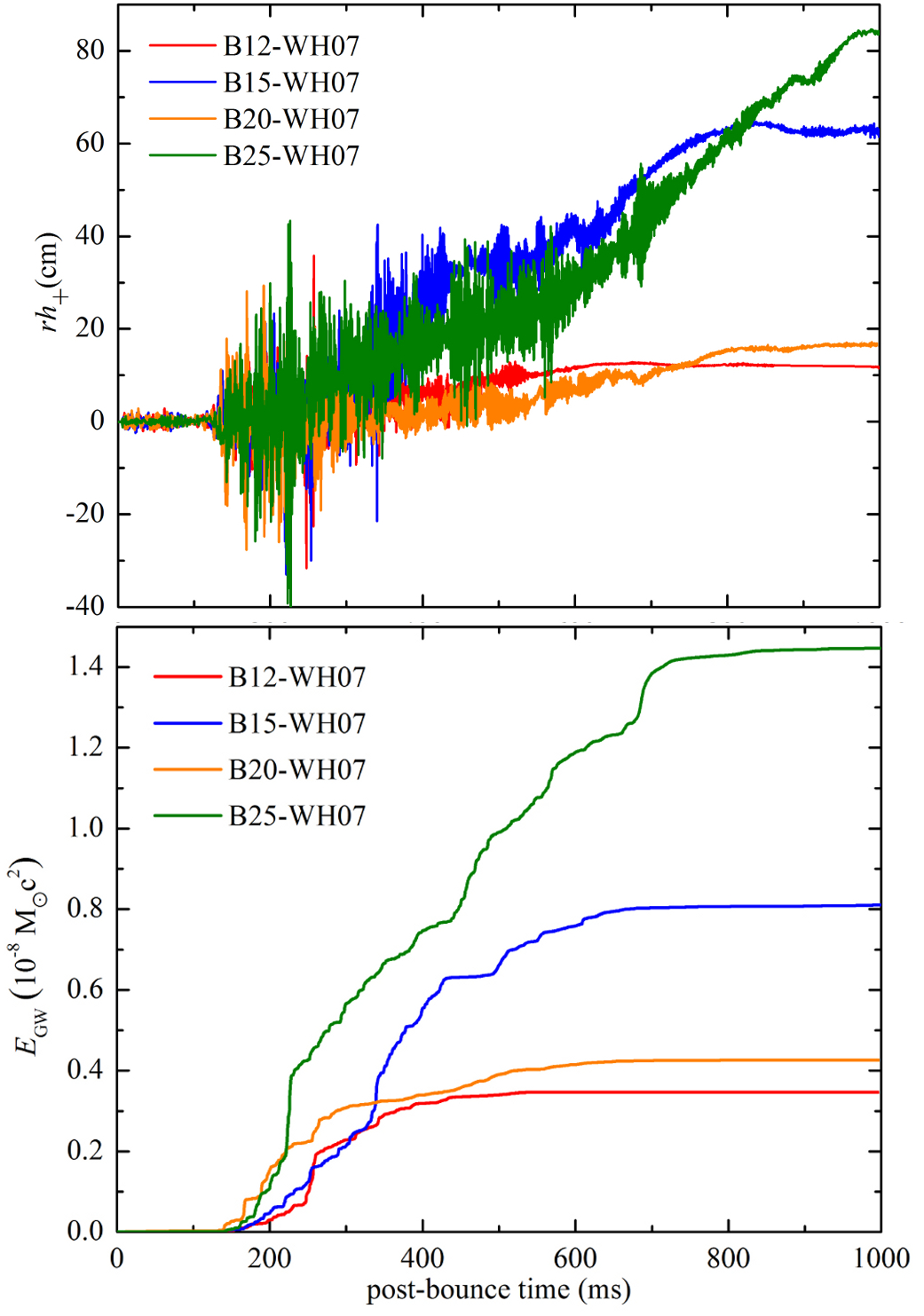}
\caption{\emph{Top:} The gravitational waveforms for all of the models presented here.
\emph{Bottom:} The energy $E_\textrm{GW}$ radiated in the form of gravitational waves as a function of time.  }
\label{fig:Egw}
\end{figure}

The total emitted GW energy is shown in Figure~\ref{fig:Egw}. The values 
of the GW energy emitted in the B-series models presented here are very close to what we predicted in the A-series models presented in \cite{Yakunin10}.
Due to the \textquotedblleft anomalous\textquotedblright evolution of model B20-WH07,  we do not observe a simple correlation between the progenitor mass and the total 
energy emitted in gravitational waves. That is,
the GW energy emitted is not solely dependent on the progenitor mass. It is also a function of the explosion dynamics -- in particular, the number and characteristics of the accretion streams that form during the pre-explosion and explosion phases. The GW energy emitted {\em does} increase monotonically with progenitor mass for models B12-WH07, B15-WH07, and B25-WH07, which share the same explosion morphology, but the results are very different for B20-WH07 given the difference in its explosion morphology and the resultant difference in the evolution of its accretion streams.
%
%

As shown in Figure~\ref{fig:Egw}, almost all of the GW energy 
is emitted between 200 ms and 700 ms after bounce for all of our models, and 
we do not observe significant contributions to $E_\textrm{GW}$ in any model after 700 ms, 
when the low-frequency component of the GW signal becomes dominant.  The jumps in the 
emitted gravitational wave energy correspond to abrupt increases in the accretion rate onto 
the PNS surface. This is easily seen by correlating these jumps with the counterpart spikes in the 
gravitational waveforms in Figure~\ref{fig:Egw} that are the direct result of such accretion.
Our predictions for the gravitational wave energy emitted are 2--3 times higher than 
those based on the general relativistic models G11.2 and G15 of  \citet{Mueller13}. 
This is largely due to the fact that the high-amplitude, GW tail contributes significantly
to the GW energy emitted. In turn, such large-amplitude tails are not produced in models
that do not explode robustly.

\subsection{Signals Produced by Neutrino Emission}

The GW signals from neutrino emission in the B-series models are similar to the signals in their A-series counterparts, 
with some variation associated with the nonlinear stochastic nature of multidimensional supernova models. 
The amplitudes of all of the GW signals from neutrino emission are slightly negative just after bounce,
until $\sim$180--220~ms, depending on the model, and then increase dramatically, becoming positive
throughout the rest of the simulation. 
This is the result of the formation of a stable accretion down flow in the central region of the grid ($60^\circ < \theta < 120^\circ$) due to an active interplay between neutrino-driven convection and the SASI at this time. The cold dense matter in the formed down flow absorbs neutrinos more efficiently than the matter in the polar regions. As a consequence, the neutrino luminosity is more intense in the polar regions, which makes the amplitude of the GW signal positive (Eq.~\ref{psi_function_eq}). The situation is more intricate in B20-WH07, due to the presence of multiple down streams, but the general trend is similar to the other models.
Note that the amplitude
of the neutrino-generated GW signal is much larger than the amplitude of the matter-generated GW
signal; however, neutrino-induced GW signals have relatively low frequencies ($f < 20$~Hz) for the 
canonical setup of current gravitational wave detectors. Nevertheless these signals may be detectable using the ``non-traditional'' approach presented in \cite{Christodoulou91, Bieri13}: three-mass experiments may allow one to measure a permanent displacement of the test masses due to the (linear) memory effect of gravitational waves.

\subsection{Detectability of the Signals}

\begin{figure}[ht!]
\centering
\includegraphics[width=86mm]{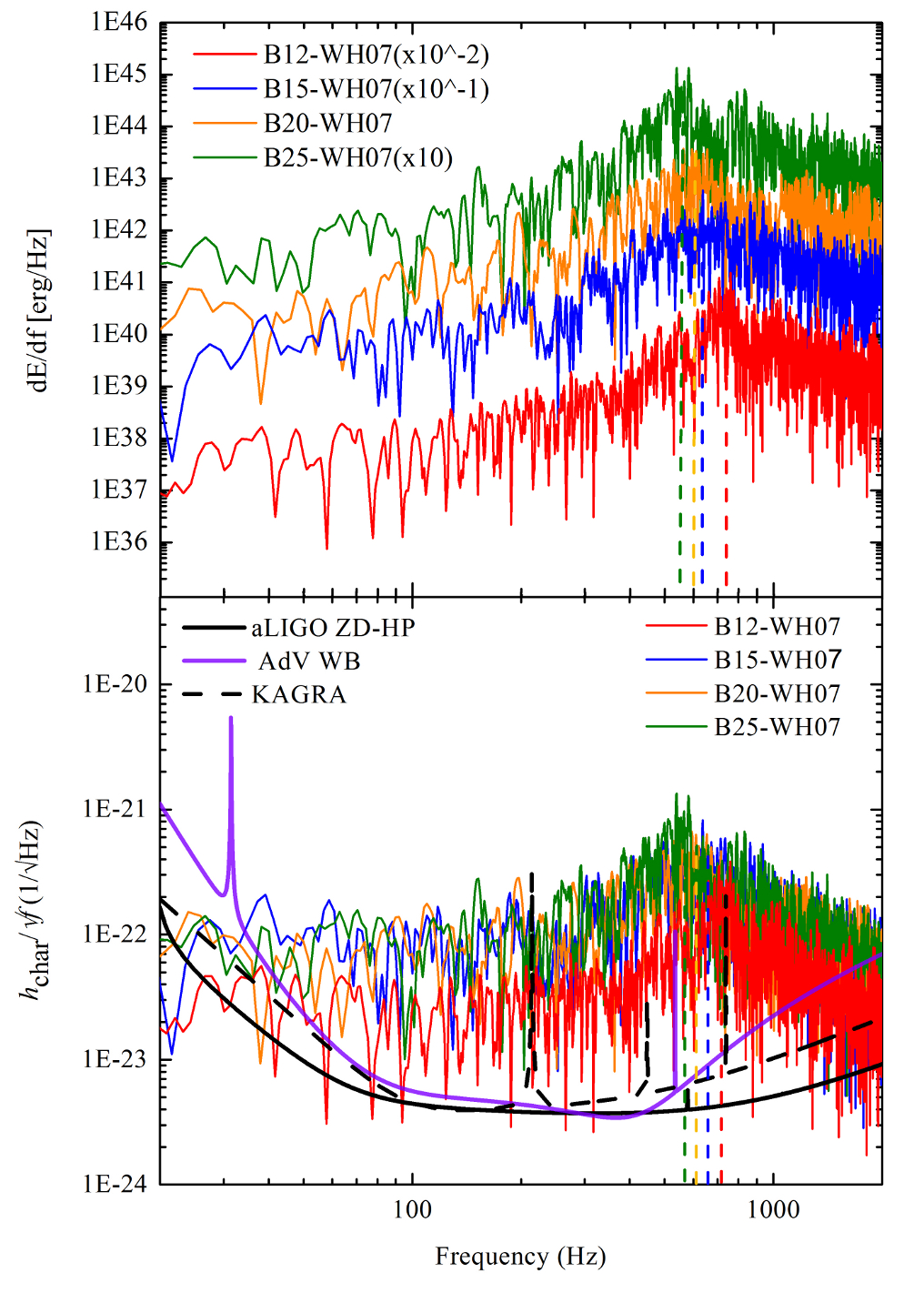}
\caption{ {\it Top:} The GW energy spectra for all of our B-series models. In order to better differentiate the curves, the spectra have been rescaled by a factor of $10^{-2}$, $10^{-1}$, and $10$ for models B12-WH07, B15-WH07, and B25-WH07, respectively.\\
{\it Bottom:} The characteristic GW amplitudes, $h_\textrm{char}$, for all of our B-series models, plotted against the approximate noise thresholds for Advanced LIGO, Advanced Virgo, and KAGRA, at a source location of 10~kpc.\\
Dashed, colored lines mark the peak frequencies of the corresponding models on both spectrographs.}
\label{Detectability}
\end{figure}

Figure~\ref{Detectability} compares the GW strain spectra $h_\textrm{char}(f)$  of our models with the broadband design noise levels of advanced-generation GW interferometers, assuming a source distance of 10~kpc. Most of the detectable emission is within $\sim$100--800~Hz, with the level increasing from ($\sim$2  to 10)$\times 10^{-23}$~Hz$^{-1/2}$. A Galactic event (at 10~kpc) appears to be well detectable by the upcoming generation of detectors. 
The peaks at $\sim$550--750~Hz are due to a cumulative effect of high-frequency convection inside the PNS and the deceleration of downflows at the PNS surface. In general, the peak frequencies of all of our models presented here are lower than those seen in the A-series models because of the low-frequency contribution of the late signal ( $>$600~ms).  
Though all of the peaks lie in a relatively narrow frequency interval, one can see that the peak frequency tends to decrease with increasing progenitor mass. 

Our GW predictions for the B15-WH07 model can be compared to the M15 model of \citet{Mueller13}
given that both groups implement similar treatments of the neutrino transport and
GR corrections to the gravitational field, and 
include essentially the same physics in their models. 
The two groups are in agreement with regard to the time scales
of the different (pre-explosion) GW phases, and the amplitude of the prompt and strong GW signals. 
They differ, however, in their predictions for the peak in the GW spectrum, which is at 
$\sim$1000~Hz in the M15 model versus 650~Hz in model B15-WH07. This
difference likely arises due to the presence of a strong explosion in our simulation that
considerably decreases active accretion onto the PNS after 400~ms post bounce and, consequently, 
the peak frequency of the GW signal (Figure~\ref{Detectability}).
\section{Summary and Conclusions}
Based on four {\em ab initio} axisymmetric explosion models \cite{Bruenn14} for non-rotating progenitors with masses 12, 15, 20, and 25~M$_\odot$, we studied the GW signals of core collapse supernovae from the early post-bounce phase through the fully-developed explosion phase. Unlike in our earlier studies \cite{Yakunin10}, which were truncated at $\sim$500 ms after bounce, here we provide the complete (up to 1 s) gravitational waveforms for all four models. This is the first time {\em complete} signals have been computed in the context of {\em ab initio} explosion models. This is particularly important for the signatures in frequency space, whose accurate determination requires a full temporal evolution.

Our models qualitatively confirm the four-phase picture of GW emission seen in our previous studies, and by others using parameterized models, with an early quasi-periodic signal, a quiescent phase of several tens of milliseconds, a strong stochastic GW signal lasting until some fraction of a second after the onset of explosion, and a low-frequency tail.

Given that we tracked the full dynamical and GW evolution beyond 1 s after bounce, we were able to follow the transitions between the four stages of the GW signal and, in particular, between the last two stages, at the moment of cessation of active accretion onto the PNS surface. This is clearly seen in the gravitational waveforms, both in the behavior of the strain as a function of time and in the evolution of the peak frequency of the GW signal, and is especially manifest in the B12-WH07 model. The evolution of the peak frequency of the signal, which declines monotonically after accretion has stopped, is clearly evident in all four cases. Moreover, the peak frequency in the GW energy spectrum and the characteristic strain is inversely related to progenitor mass. On the other hand, the total GW energy emitted does not exhibit a simple correlation with progenitor mass. For models B12-WH07, B15-WH07, and B25-WH07, which all exhibit the same prolate explosion morphology, the GW energy emitted increases monotonically with progenitor mass. However, model B20-WH07, with its more spherical explosion, emits significantly less energy in GWs than models B25-WH07 and B15-WH07, which indicates that the energy emitted in GWs is a gross function of two parameters: progenitor mass and explosion morphology.

Our findings clearly demonstrate that a complete prediction of gravitational waveforms and spectra is possible only for a \emph{ab initio} simulation with a fully developed explosion. Both the peak frequency of the GW signal and the GW energy emitted depend on the duration of the GW signal. In the latter case, the GW energy emitted does not saturate until 400--700 ms after bounce, depending on the progenitor.

In addition to carrying out our simulations past 1 s after bounce, differences between the signals provided here and those published in our earlier work \cite{Yakunin10} were obtained for the 
early, first-phase signal and, as discussed here, were the result of different model specifications -- in particular, constraining the pre-shock flow to be spherically symmetric or not. The results described here were obtained in models that do not impose such a constraint, which is artificial.

A major shortcoming of the work presented here is our imposition of axisymmetry. Complete GW waveform predictions, including both the $h_+$ and $h_\times$ polarizations, in the context of state-of-the-art three-dimensional supernova simulations are required. The expected differences in the stellar core hydrodynamics in three dimensions case \cite{Nordhaus10, Burrows12, Janka12, Hanke13, Murphy13, Dolence13, Couch13, Ott13, Takiwaki14, Fuller15, Radice15, Couch15} will have an impact on the predicted waveforms. In particular, changes in the geometry of the accretion down flows and the behavior of turbulence may influence the amplitude and energy of the GW emission.
We have already reported a successful development of supernova explosion in the 3D \chimera simulation \cite{Mezzacappa15}. Within a month, we will present an overview of the simulation \cite{Lentz15} followed by the paper 	presented the 3D gravitational waveforms. 

In addition to the leap to three dimensions, other \textquotedblleft dimensions\textquotedblright~of the problem -- e.g., the use of different nuclear equations of state, progenitors (especially non-spherical progenitors (see \cite{ArMe11}), etc. -- need to be considered, as well, in the context of {\em late-time} three-dimensional models.
%
\\
\par
This research was supported by the U.S. Department of Energy Offices of Nuclear Physics and Advanced Scientific Computing Research; the NASA Astrophysics Theory and Fundamental Physics Program (grants NNH08AH71I and NNH11AQ72I); and the National Science Foundation PetaApps Program (grants OCI-0749242, OCI-0749204, and OCI-0749248). 
P. M. is supported by the National Science Foundation through its employee IR/D program. The opinions and conclusions expressed herein are those of the authors and do not represent the National Science Foundation. The authors would like to acknowledge fruitful discussions with Nelson Christensen.
  
\bibliographystyle{apsrev4-1} 

\bibliography{GW-B_main} 
%
%
\appendix

\newcommand{\tr}{\mbox{tr}}

\section{Tensor Spherical Harmonics $f_{lm}$}

In this appendix, we write explicitly the tensor spherical harmonics.
We follow the notation of \cite{Kotake07}.

\subsection{Explicit Form of $W_{lm}$ and $X_{lm}$}

In general, the $W_{lm}$ and $X_{lm}$ functions are defined by

\begin{eqnarray}
 W_{lm}(\theta,\phi) &=& \left(\frac{\partial^2}{\partial\theta^2}
-\cot\theta\frac{\partial}{\partial\theta}-\frac{1}{\sin^2\theta}
\frac{\partial^2}{\partial\phi^2}\right) 
Y_{lm} (\theta,\phi)
\\
 X_{lm}(\theta,\phi) &=& 2\frac{\partial}{\partial\varphi}
\left(\frac{\partial}{\partial\theta}-\cot\theta\right)
Y_{lm}(\theta,\phi),
\end{eqnarray}
where $Y_{lm}$ is the usual scalar spherical harmonics of order $l$ and degree $m$. 

For $l=2$ (quadrupole approximation), we have the following explicit expressions:

\begin{eqnarray}
 X_{2,2} &=& i\sqrt{\frac{15}{2\pi}}\sin\theta\cos\theta e^{2i\phi},\\
 X_{2,1} &=& i\sqrt{\frac{15}{2\pi}}\sin^2\theta e^{i\phi},\\
 X_{2,0} &=& 0, \\
 X_{2,-1} &=& i\sqrt{\frac{15}{2\pi}}\sin^2\theta e^{-i\phi},\\
 X_{2,-2} &=& -i\sqrt{\frac{15}{2\pi}}\sin\theta\cos\theta e^{-2i\phi},
\end{eqnarray}
and
\begin{eqnarray}
 W_{2,2} &=& \sqrt{\frac{15}{2\pi}}\frac{1+\cos^2\theta}{2}e^{2i\phi},\\
 W_{2,1} &=& \sqrt{\frac{15}{2\pi}}\sin\theta\cos\theta e^{i\phi},\\
 W_{2,0} &=& \sqrt{\frac{15}{4\pi}}3\sin^2\theta,\\
 W_{2,-1} &=& -\sqrt{\frac{15}{2\pi}}\sin\theta\cos\theta e^{-i\phi},\\
 W_{2,-2} &=& \sqrt{\frac{15}{2\pi}}\frac{1+\cos^2\theta}{2}e^{-2i\phi}.
\end{eqnarray}

\subsection{Tensor Spherical Harmonics}

Let us define the tensor spherical harmonics (in the $O$-frame) using
the $W_{lm}$ and $X_{lm}$ functions
\begin{equation}
 f_{l m} = \alpha
  \begin{bmatrix}
   W_{l m} & X_{l m} \\
   X_{l m} & -W_{l m}\sin^2\theta
  \end{bmatrix}.
\end{equation}
Here the first row (column) corresponds to $\theta$ ($\phi$),
and $\alpha$ is a normalization
factor. Notice, this tensor is trace-free, and the diagonal components correspond to the $+$ mode, while the off-diagonal components 
correspond to the $\times$ mode.

The normalization is fixed by the following relation:

\begin{equation}
 \int d\Omega \left(f_{l m}\right)_{AB}
\left(f_{l' m'}^*\right)_{CD}
({}^2\gamma)^{AC}({}^2\gamma)^{BD}
= \delta_{l l'}\delta_{m m'},
\end{equation}
where $A,B,C,D=\theta,\phi$, and the metric on the 2-sphere,
${}^2\gamma_{AB}$, is 

\begin{equation}
 {}^2\gamma_{AB} = 
  \begin{bmatrix}
   1 & 0 \\
   0 & \sin^2\theta
  \end{bmatrix}.
\end{equation}
The normalization factor for $l=2$ is $\frac{1}{4\sqrt{3}}$.

\end{document}